\documentclass[aps,prd,superscriptaddress,nofootinbib]{revtex4}

\usepackage{graphicx}

\usepackage{amsmath}

\usepackage{amssymb}

\usepackage{pstricks}

\begin{document}

\title{Exploring the dark side of the Universe in a dilatonic brane-world scenario\footnote{Contribution to the anniversary volume ``\textit{The Problems of Modern Cosmology}'', on the occasion of the 50th birthday of \mbox{Prof. Sergei D. Odintsov}. Editor: Prof. P. M. Lavrov, Tomsk State Pedagogical University.}}

\author{Mariam Bouhmadi-L\'{o}pez}
\email{mariam.bouhmadi@fisica.ist.utl.pt}
\affiliation{Centro Multidisciplinar de Astrof\'{\i}sica - CENTRA, Departamento de F\'{\i}sica, Instituto Superior T\'ecnico, Av. Rovisco Pais 1,
1049-001 Lisboa, Portugal}

\begin{abstract}

We describe the late-time acceleration of the Universe within the paradigm of the brane-world scenario. More precisely, we show how a phantom-like behaviour or a crossing of the cosmological constant line can be achieved safely in a dilatonic brane-world model with an induced gravity term on the brane. The brane tension plays the role of dark energy which is coupled to the dilaton bulk scalar field. The phantom mimicry  as well as the crossing of the cosmological constant line
are achieved without invoking any phantom matter either on the brane or in the bulk. 

\end{abstract}

\date{\today}

\maketitle

\section{Introduction}\label{sec1}

Understanding the recent acceleration of the universe is a challenge and a landmark problem in physics. Its resolution may affect in the  short term our understanding of a fundamental interaction like gravity as well as enlarge the framework of particle physics. The smoking gun of the acceleration of the universe (if we assume it is homogeneous and isotropic on large scales) was provided  by the analysis  of the Hubble diagram of SNe Ia a decade ago \cite{Perlmutter:1998np}. This discovery, together with (i) the measurement of the fluctuations in the CMB which implied that the universe is (quasi)  spatially flat and (ii) that the amount of matter which clusters gravitationally is much less than the critical energy density, implied the existence of a ``dark energy component'' that drives the late-time acceleration of the universe. Subsequent precision measurements of the CMB anisotropy by WMAP \cite{Spergel:2003cb} and the power spectrum of galaxy clustering by the 2dFGRS and SDSS  surveys \cite{Cole:2005sx,Tegmark:2003uf} have confirmed this discovery.

A plethora of different theoretical models have been so far proposed to explain this phenomenon \cite{Copeland}, although unfortunately none of the models advanced so far is both completely convincing and well motivated. A cosmological constant corresponding to roughly two thirds of the total energy density of the universe is perhaps the simplest  {\it phenomenological} way  to explain the late-time speed up of the universe --and in addition   match rather well the observational data. However,  the expected theoretical value of the cosmological constant is about 120 orders of magnitude larger than the value needed to fit the data \cite{Durrer:2007re}. 

Alternative approaches to explain the late-time acceleration invoke (i) a dark energy component in the universe which would provide a negative pressure or (ii) an  infrared modification of general relativity  on large scales (like in some brane-world scenarios \cite{Dvali:2000hr} or f(R) models \cite{frmodels}) which, by weakening the gravitational interaction on those scales, allows the recent speed up of the universal expansion. The second approach is also motivated by the fact that we only have precise measurements of gravity from sub-millimiter scales up to solar system scales while the Hubble radius, which is the scale relevant for the cosmic acceleration, is many orders of magnitude larger.

A pioneering scheme in the second approach is the Dvali, Gabadadze and Porrati (DGP) model \cite{Dvali:2000hr} which  corresponds to a 5-dimensional (5D) induced gravity brane-world model \cite{Deffayet,IG, Sahni:2002dx,LDGP2,Bouhmadi-Lopez:2004ys}, where a low-energy modification occurs with respect to general relativity; i.e. an infrared  effect takes place, leading to two  branches of solutions: (i) the self-accelerating branch and (ii)  the normal branch.

The self-accelerating branch solution gives
rise to an asymptotically de Sitter brane; i.e. a late-time accelerating brane universe. The acceleration of
the brane expansion arises naturally, i.e. without invoking the
presence of any dark energy on the brane to produce the speed-up.
Most importantly, it has recently been shown  that by
embedding the DGP model in a higher dimensional space-time the ghost
problem in the original model \cite{Koyama:2007za} may be cured  \cite{deRham:2007xp} while preserving the existence of a self-accelerating solution \cite{Minamitsuji:2008fz}. 

The normal branch also constitutes in itself an extremely interesting physical setup
of the DGP model however, as it can mimic a phantom behaviour on
the brane by means of the $\Lambda$DGP scenario
\cite{Sahni:2002dx}. We would like to highlight that
observational data do not seem incompatible with a phantom-like
behaviour \cite{Percival:2007yw} and therefore we should keep an open
mind about what is producing the recent inflationary era of our
universe. Furthermore, this phantom-like behaviour may well be a
property acquired only recently by dark energy. This leads to an
interest in modelling  the so called crossing of the
phantom divide line $w=-1$; for example in the context of the
brane-world scenario \cite{LDGP2,crossing,BouhmadiLopez:2008bk}. The most important aspect
of the $\Lambda$DGP model is that the phantom-like mimicry is obtained
without invoking any real phantom-matter  \cite{phantom} which is
known to violate the null energy condition and induce quantum
instabilities\footnote{We are referring here to a phantom energy
  component described through a minimally coupled scalar field with
  the wrong kinetic term.} \cite{Cline:2003gs}. In the  DGP  scenario
it is as well  possible to get a mimicry of the crossing of the phantom divide, however,  at the cost of invoking a dynamical dark energy on the brane \cite{LDGP2}, for example modelled by a quiessence fluid or a (generalised)  Chaplygin gas.

One aim of this paper is to show that a dilatonic brane-world model with an induced gravity term in the brane can mimic a phantom-like behaviour without including matter on the brane that violates the null  energy condition. A second aim of this paper is to  show that there is an alternative form  (to the one introduced in \cite{Sahni:2002dx,LDGP2}) of mimicking the crossing the cosmological constant line $w=-1$ in the brane-world scenario. 
More precisely, we consider a 5D dilatonic bulk with a brane endowed with an induced gravity term, a brane matter content corresponding   to cold dark matter, and a brane tension $\lambda$ that depends on the minimally coupled bulk scalar field. We will show that in this set-up the vacuum generalised self-accelerating branch expands in a super-accelerating way and mimics a  phantom-behaviour. On the other hand, the generalised normal branch expands in an accelerated way due to $\lambda$ playing the role of \textit{dark energy} --through its dependence on the bulk scalar field. Furthermore, in this case, it turns out that  the brane tension grows with the brane scale factor until it reaches a maximum positive value and then starts decreasing. Therefore, in our model the brane tension mimics a crossing of the phantom divide. Most importantly no matter violating  the null energy density is invoked in our model. 

The paper is organised as follows. In section 2, we present our dilatonic brane-world model with induced gravity. The bulk scalar field potential is an exponential  potential. The matter content of the brane is coupled to the dilaton field. We deduce the modified Friedmann equation for both branches, the junction condition of the dilaton across the brane, which constrains the brane tension, and the energy balance on the brane. In section 3, we then analyse the vacuum (i.e., $\rho_{m}=0$) solutions in both branches. We show that the brane tension has a phantom-like behaviour on the generalised self-accelerating branch in the sense that the brane tension grows as the brane expands. In this branch, the brane hits a singularity in its future evolution which may be interpreted as a ``big rip'' singularity pushed towards an infinite cosmic time. Then, in section 4, we show that, under some assumptions on
the nature of the coupling parameters between $\lambda$ and $\phi$,
$1+w_{\rm{eff}}$ changes sign as the normal brane evolves, with $w_{\rm{eff}}$ the
effective equation of state for the brane tension. Our
conclusions are presented in section 5.

\section{The framework}\label{sec2}

We consider a  brane, described by a 4D hyper-surface ($h$, metric g), embedded in a 5D bulk space-time ($\mathcal{B}$, metric 
$g^{(5)}$), whose action is given by 
%%%%%%%%%%%%%
\begin{eqnarray}
\mathcal{S} = \,\,\, \frac{1}{\kappa_5^2}\int_{\mathcal{B}} d^5X\, \sqrt{-g^{(5)}}\;
\left\{\frac{1}{2}R[g^{(5)}]\;+\;\mathcal{L}_5\right\}%\nonumber \\
 + \int_{h} d^4X\, \sqrt{-g}\; \left\{\frac{1}{\kappa_5^2} K\;+\;\mathcal{L}_4 \right\}\,, \label{action}
\end{eqnarray}
%%%%%%%%%%%%%%%
where $\kappa_5^2$ is the 5D gravitational constant,
$R[g^{(5)}]$ is the scalar curvature in the bulk and $K$ the extrinsic curvature of the brane in the higher dimensional
bulk, corresponding to the York-Gibbons-Hawking boundary term.  

We consider a dilaton  field $\phi$ living on the bulk \cite{Chamblin:1999ya,Maeda:2000wr} and we choose $\phi$ to be dimensionless. 
Then, the 5D Lagrangian $\mathcal{L}_5$ can be written as
%%%%%%%%%%%%%%%
\begin{eqnarray} 
\mathcal{L}_5=-\frac12 (\nabla\phi)^2 -V(\phi).
\end{eqnarray}

The 4D Lagrangian $\mathcal{L}_4$ corresponds to 
%%%%%%%%%%%%%%%%
\begin{equation}
\mathcal{L}_4= \alpha {R}[g] -\lambda(\phi)+\Omega^{4}\mathcal{L}_m(\Omega^2 g_{\mu\nu}).%e^{4b\phi}\mathcal{L}_m(e^{2b\phi}g_{\mu\nu}),
\label{L4}\end{equation}
%%%%%%%%%%%%%%%%
The first term on the right hand side (rhs) of the previous equation corresponds to an induced gravity term \cite{Dvali:2000hr,Deffayet,IG, Sahni:2002dx},
where $R[g]$ is the scalar curvature of the induced metric on the
brane and $\alpha$ is  a positive parameter which  measures the strength of the induced gravity term 
and has dimensions of mass squared. The term $\mathcal{L}_m$ in Eq.~(\ref{L4}) describes  the matter content of the brane and $\lambda(\phi)$ is the brane tension, and we will restrict ourselves to the case where they are homogeneous and isotropic on the brane.
We allow the brane matter content to  be  non-minimally coupled on the (5D) Einstein frame  but to be minimally coupled respect to a conformal metric 
${\tilde g}^{(5)}_{AB}=\Omega^2\;g^{(5)}_{AB}$, where $\Omega=\Omega(\phi)$ \cite{Maeda:2000wr}.

We are interested in the cosmology of this model. It is known that
for an expanding FLRW brane the unique bulk space-time in Einstein gravity (in vacuum) is a 5D Schwarzschild-anti de Sitter space-time. This property as far as we know cannot be extended  to a 5D dilatonic bulk. On the other hand, the presence of an induced gravity term in the  brane-world scenario affects only the dynamics of the brane, through the junction conditions at the brane, and does not affect the bulk field equations. Therefore, in order to study the  effect of an induced gravity term in a brane-world dilatonic model, it is possible to consider a bulk corresponding to a dilatonic 5D space-time and later on impose the junction conditions at the brane location. The junction conditions will then determine the cosmological evolution of the brane and constrain the brane tension. This is the approach we will follow. 

From now on, we consider a 5D dilatonic solution obtained by Feinstein et al \cite{Feinstein:2001xs,Kunze:2001ji}
\textit{without an induced gravity term on the brane}. The 5D dilatonic solution reads \cite{Kunze:2001ji}
\begin{equation}
ds^2_5=\frac{1}{\xi^2}r^{2/3(k^2-3)}dr^2 +r^2(-d{t}^2+\gamma_{ij}dx^idx^j), 
\label{bulkmetric}\end{equation}
where  $\gamma_{ij}$ is a 3D spatially flat  metric. The bulk potential is
\begin{equation}
V(\phi)=\Lambda\exp[-(2/3) k\phi].
\label{liouville}\end{equation}
The parameters $k$ and $\xi$ in Eq.~(\ref{bulkmetric}) define the 5D cosmological constant $\Lambda$ 
\begin{equation} \Lambda= \frac12 (k^{2}-12)\xi^2. \end{equation}
The 5D scalar field grows logarithmically with the radial coordinate $r$ \cite{Kunze:2001ji}
\begin{equation} \phi=k\log (r). \label{phi}\end{equation}

Now, we consider a FLRW brane filled only with cold dark matter (CDM); i.e pressureless matter,  and the brane tension $\lambda(\phi)$.  On the other hand, the brane is considered to be embedded in the previous 5D dilatonic solution  
and its trajectory in the bulk is described by the following parametrisation 
\begin{equation}{t}={t}(\tau),\,\,\,\, r=a(\tau),\,\,\,\, x_i= constant,\,\, i=1 \ldots 3. \end{equation}
Here $\tau$ corresponds to the brane proper time. Then the brane metric reads
\begin{equation} 
ds^2_4\,=\,g_{\mu\nu}\,dx^{\mu}dx^{\nu}\,=\,-d\tau^2+a^2(\tau)\gamma_{ij}dx^idx^j.
\end{equation}
For an induced gravity brane-world model
\cite{Deffayet,Bouhmadi-Lopez:2004ys}, there are two physical ways of
embedding the brane in the bulk when a $\mathbb{Z}_2$-symmetry across
the brane is assumed: the generalised normal branch\footnote{We will refer to the normal DGP branch also as the non-self-accelerating DGP branch.} and the generalised self-accelerating branch. For example, in the first case the brane is moving in the bulk away from the bulk naked singularity located at $r=0$ \cite{Bouhmadi-Lopez:2004ys}.

For simplicity, we will consider that the matter content of the brane is minimally coupled respect to the  conformal metric ${\tilde g}^{(5)}_{AB}=\exp(2b\phi)\;g^{(5)}_{AB}$; i.e. $\Omega=\exp(b\phi)$, where $b$ is a constant.
We will also consider only the case\footnote{The main conclusions  do not depend on the sign of $k$ but on the sign of the parameter $kb$. Therefore, we can always describe the same physical situation on the brane for $k<0$ by changing the sign of $b$.} $k > 0$; i.e. the scalar field is a growing function of the coordinate $r$.
Then, the Israel junction condition at the brane \cite{Chamblin:1999ya} describes the cosmological evolution of the brane through the modified Friedmann equation, which in our case reads
\begin{equation} \sqrt{{\xi^2}{a^{-\frac23 k^2}}+H^2} =-\epsilon\,\frac{\kappa_5^2}{6}
\left[\lambda(\phi)+\rho_m-6\alpha H^2\right],
\label{Friedmann1}\end{equation}
where $\epsilon=1$ for the self-accelerating brane and  $\epsilon=-1$ for the normal branch. The modified Friedmann equation can be  more conveniently expressed as
\begin{eqnarray} 
H^2=%&&
\frac{1}{6\alpha}\left\{\lambda+\rho_m%\right.\label{Friedmann}\\
%&&\hspace*{-0.9cm}+\left. 
+\frac{3}{\kappa_5^4\alpha}
\left[1+\epsilon \sqrt{1+4\kappa_5^4\alpha^2\xi^2a^{-2k^2/3}+\frac{2}{3}\kappa_5^4\alpha(\lambda+\rho_m)}\right]\right\},%\nonumber
\label{Friedmann}
\end{eqnarray}
where $\lambda$ is the brane tension and $\rho_m$ is the energy density of CDM.

On the other hand, as it is usual in a dilatonic brane-world scenario, matter on the brane --in this case CDM-- is not conserved due to the coupling $\Omega$ (see Eq.~(\ref{L4})). In fact, we have
\begin{equation}
\dot\rho_m=-3H\left(1-\frac13 kb\right)\rho_m, 
\label{conservationrho}
\end{equation}
where a dot stands for derivatives respect to the brane cosmic time $\tau$. Therefore, CDM on the brane scales  as
\begin{equation}
\rho_m=\rho_0 a^{-3+kb}.
\label{dust}
\end{equation}

Finally, the junction condition of the scalar field at the brane \cite{Chamblin:1999ya} constrains the brane tension $\lambda(\phi)$. In our model this is given by
\begin{equation}
a\frac{d\lambda}{da}=-kb\rho_m\, {+\epsilon}\frac{2k^2}{\kappa_5^2}\sqrt{{\xi^2}{a^{-\frac23 k^2}}+H^2},
\label{constraint}
\end{equation}
where for convenience we have rewritten the scalar field (valued at the brane) in terms of the scale factor of the brane. At this respect we remind the reader that  at the brane $\phi=k\log(a)$. 

\section{Vacuum solutions}\label{sec3}

The vacuum solutions, i.e. in absence of matter on the brane, depends crucially on the embedding of the brane in the bulk, therefore, which branch we are considering.

\subsection{The self-accelerating branch}

For the vacuum self-accelerating branch; i.e.  $\epsilon=1$ and $\rho_m=0$, the brane tension is an  increasing function  of the scale factor of the brane\footnote{The constraint equation (\ref{constraint}) (after substituting the Hubble rate given in Eq.~(\ref{Friedmann})) can be solved analytically in this case \cite{Bouhmadi-Lopez:2004ys} and it can be explicitly shown that the brane tension increases as the brane expands. In  the same way a parametric  expression can be found for the Hubble rate and its cosmic derivative. \label{footnote3new}} (see Eq.~(\ref{constraint})). For small values of the scale factor, the brane tension reaches infinite negative values. On the other hand, for very large value of the scale factor the brane tension approaches infinite positive values. Therefore, when the brane tension acquires positive values, it mimics a phantom energy component in a standard FLRW universe. We remind at this respect that we have not included any matter that violates the null energy condition; i.e. any explicit phantom energy in the model.

The Hubble parameter is  an increasing function of the scale factor, i.e the brane super-accelerates. In fact, 
\begin{equation}
\dot H=-\frac{ k^2 H^2}{\kappa_5^4\alpha(\lambda-6\alpha H^2)+3}, \label{Rayeq}
\end{equation}
while the modified Friedmann equation (\ref{Friedmann}) implies that the denominator of the previous equation has to be negative (see also footnote \ref{footnote3new}), consequently $\dot H >0$. At small scale factors, $H$ reaches  a constant positive value. Therefore, in the vacuum self-accelerating brane there is no big bang singularity on the brane; indeed, the brane is asymptotically de Sitter. On the other hand, at very large values of the scale factor, the Hubble parameter diverges. 

The divergence of the Hubble parameter for very large values of the scale factor  might point out the existence of a big rip singularity in the future evolution of the brane; i.e. the scale factor and Hubble parameter blow up in a finite cosmic time in the future evolution of the brane. However, it can be shown that the divergence of $H$  and  $a$ (and also of $\lambda$) occur in an infinite cosmic time in the future evolution of this branch. This can be easily proven by noticing that the asymptotic behaviour of the Hubble parameter  at large value of the scale factor is  
\begin{equation}
H\sim\frac{k^2}{\kappa_5^2 \alpha}\ln (a).
\end{equation}
Consequently, the Hubble rate does not grow as fast as in phantom energy models with a constant equation of state where a big rip singularity takes place on the future evolution of a homogeneous and isotropic universe \cite{phantom}. 

In summary, we have proven that in the vacuum self-accelerating branch the brane tension mimics a phantom behaviour. On the other hand,  there is a singularity in the future evolution of the brane. The singularity is  such that for large value of the cosmic time, the scale factor and the Hubble parameter diverge. This kind of singularity can be interpreted  as a ``big rip''  singularity pushed towards an infinite cosmic time of the brane. 

\subsection{The normal branch}

For the vacuum normal branch; i.e.  $\epsilon=-1$ and $\rho_m=0$, the brane tension is a decreasing function  of the scale factor of the brane\footnote{The constraint equation (\ref{constraint}) (after substituting the Hubble rate given in Eq.~(\ref{Friedmann})) can be solved analytically in this case \cite{Bouhmadi-Lopez:2004ys} and it can be explicitly shown that the brane tension decreases as the brane expands. In  the same way a parametric  expression can be found for the Hubble rate and its cosmic derivative. \label{footnote3}} (see Eq.~(\ref{constraint})). For small values of the scale factor, the brane tension reaches infinite positive values. On the other hand, for very large value of the scale factor the brane tension vanishes. 

The Hubble parameter is  a decreasing function of the scale factor, i.e the brane is never super-accelerating. In fact, Eq.~(\ref{Rayeq}) and the Israel junction condition (\ref{Friedmann1}) implies that the denominator of the previous equation has to be positive (see also footnote \ref{footnote3}), therefore $\dot H<0$. At high energy, $H$ reaches  a constant positive value. Consequently, in the vacuum brane there is no big bang singularity on the brane; indeed, the brane is asymptotically de Sitter. On the other hand, at very large values of the scale factor, the Hubble parameter vanishes (the brane is asymptotically Minkowski in the future). Although the brane never super-accelerates, the brane always undergoes an inflationary period. 

\begin{figure}[h]
\includegraphics[width=0.5\textwidth]{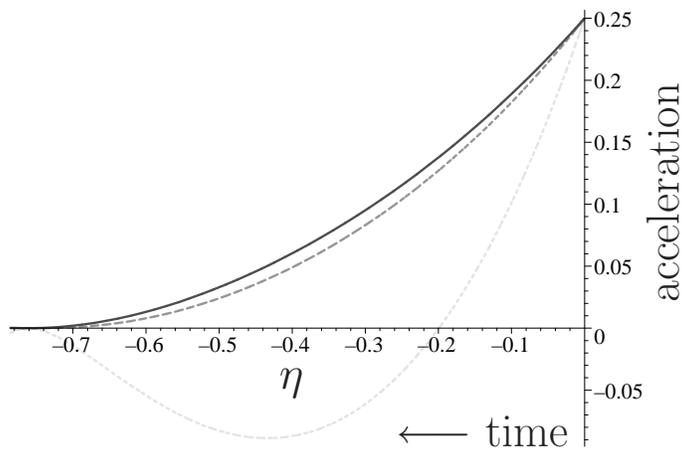}
\caption{This figure shows the behaviour of the dimensionless  acceleration parameter given by $\alpha^2\kappa_5^4\ddot{a} /a$ as a function of the time (see the left hand side arrow). The solid (darker grey), dotted and dashed-dotted (lighter grey) lines correspond to the acceleration parameter for $k^2=2,3,15$ respectively. For $k^2= 2,3$ the negative branch is eternally inflating. On the other hand, for $k^2= 15$ the brane undergoes an initial transient inflationary epoch.}
\label{inflation2}
\end{figure}

The brane behaves in two different ways depending on the value taken
by $k^2$ (cf. Fig.~\ref{inflation2}). Thus, for $k^2\leq 3$ the brane is eternally inflating. A
similar behaviour was found in \cite{Kunze:2001ji}.  On the other
hand, for $k^2>3$ the brane undergoes an initial stage of inflation
and later on it starts decelerating. This second behaviour contrasts
with the results in \cite{Kunze:2001ji} for a vacuum brane without an
induced gravity term on the brane. Then, the inclusion of an induced gravity term on a dilatonic brane-world model with an exponential potential in the bulk allows for the normal branch to inflate  in a region of parameter space where the vacuum dilatonic brane alone would not inflate. This behaviour has some similarity with steep inflation \cite{Copeland:2000hn}, where high energy corrections to the Friedmann equation in RS scenario \cite{Randall:1999vf} permit an inflationary evolution of the brane with potentials too steep to sustain it in the standard 4D case, although the inflationary scenario introduced by Copeland et al in  \cite{Copeland:2000hn} is supported by an inflaton confined in the brane while in our model inflation on the brane is induced by a dilaton field on the bulk. 

\section{Crossing the cosmological constant line}\label{sec4}

We now address the following question: is it possible to mimic a crossing of the phantom divide in particular in the model introduced in section \ref{sec2}? Unlike the vacuum case --which can be solved analytically \cite{Bouhmadi-Lopez:2004ys}-- in this case we cannot exactly solve the constraint (\ref{constraint}). Nevertheless, we can answer the previous question based in some physical and reasonable assumptions  and as well as on numerical methods. For simplicity, we will restrict to the normal branch. 

In order to answer the previous question,  it is useful to introduce the following dimensionless quantities
\begin{eqnarray}
&&\bar\lambda\equiv\frac23\kappa_5^4\alpha\lambda,\,\, x\equiv \frac23 k\phi-\ln d, \,\, d\equiv 4\alpha^2\kappa_5^4 \xi^2,\,\, m\equiv 3-kb,\nonumber\\
&&\beta_0\equiv\frac{9\beta_2}{2k^2},\,\,\beta_1\equiv\frac{2\kappa_5^4\alpha}{m}\rho_0d^{-\beta_0},\,\, \beta_2\equiv\frac{m}{3}. \label{def}
\end{eqnarray}
In terms of these new variables, the constraint given in Eq.~(\ref{constraint}) reads
\begin{eqnarray}
\frac{d\bar\lambda}{dx}=1-\beta_0\beta_1(1-\beta_2)e^{-\beta_0x}-
\sqrt{1+\bar\lambda+e^{-x}+\beta_1\beta_2 e^{-\beta_0x}}.
\label{dvdx}
\end{eqnarray}

\begin{figure}[h]
\begin{center}
\includegraphics[width=0.45\textwidth]{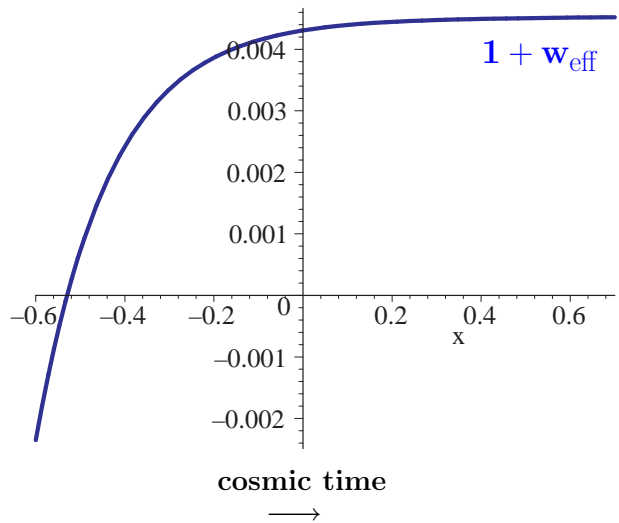}
\end{center}
\caption{The figure shows the effective equation of state of the brane tension defined in Eq.~(\ref{eqstate}) against the variable $x$ defined in Eq.~(\ref{def}). Notice that $x$ grows as the brane expands and therefore $dx/d\tau>0$ where $\tau$ corresponds to the cosmic time of the brane. This illustrative numerical solution has been obtained for $b=-1$, $k=1$ and $\beta_1=1$. The last parameter is defined in Eq.~(\ref{def}). In order to impose the right initial condition, we started the integration well in the past where CDM dominated over the scalar field on the brane and we took as a good approximated solution the dark matter solution given in Eq.~(\ref{asymlambda11}).}
\label{efectomenweff}
\end{figure}

The  assumptions we make are the following:

\begin{enumerate}
\item We assume that CDM dominates over the vacuum term ($a^{-2/3
    k^2}$) at early times on the brane. This implies that the
  parameter $\beta_0$ introduced in Eq.~(\ref{def}) must satisfy $\beta_0 > 1$. On the other hand, the brane tension will play the role of dark energy (through its dependence on the scalar field) in our model. This first assumption assumes that dark matter dominates over dark energy at high redshift which is a natural assumption to make. Indeed, at high redshift the brane tension would scale as 

\begin{equation}
\bar\lambda\sim\beta_1(1-\beta_2)e^{-\beta_0 x} + \ldots.
\label{asymlambda11}
\end{equation}

\item We also assume that CDM redshifts away a bit faster than usual; i.e. $bk<0$ or $\beta_2$ introduced in Eq.~(\ref{def}) is such that $\beta_2 > 1$. This lost energy will be used to increase the value of the scalar field $\phi(a)$ on the brane. That is, to push the brane to higher values of $a$. 
\item Finally, we also assume that $\beta_2< 2\beta_0(\beta_2-1)$. This condition, together with $\beta_0, \beta_2 >  1$, is sufficient to prove the non existence of a local minimum of the brane tension during the cosmological evolution of the brane. In fact, we can show the existence of a unique maximum for an even larger set of parameters $\beta_0>1/2$,  $\beta_2 > 1$ and $\beta_2< 2\beta_0(\beta_2-1)$. Therefore,
the set of allowed parameter $k$ and $b$ that fulfil the last three inequalities
are such that
\begin{equation}
k< {\rm min}\left\{-3b,\frac32\left[-b+\sqrt{b^2+4}\right]\right\}=-3b.
\end{equation}
where $b$ is positive.
\end{enumerate}
 
Under these three assumptions, it can be shown that the brane tension has a local maximum which must be positive (we refer the reader to \cite{BouhmadiLopez:2008bk}  for a detailed proof). In fact, what happens under these conditions is that the brane tension increases until it reaches its maximum positive value and then it starts decreasing. It is precisely at this maximum that the brane tension mimics crossing the phantom divide. Around the local maximum of the brane tension we can always define an effective equation of state in analogy with the standard 4D relativistic case:

\begin{equation}
1+w_{\textrm{eff}}=-\frac{1}{3H}\frac{1}{\lambda}\frac{d\lambda}{d\tau}.
\label{eqstate}
\end{equation}
As we mentioned earlier, the constraint equation (\ref{constraint}) cannot be solved analytically and therefore we have to resort to numerical methods. We show in Fig.~\ref{efectomenweff} an example of  our numerical results where it can be seen clearly that $1+w_{\textrm{eff}}$ changes sign. It is precisely at that moment that the crossing takes place.

\begin{figure}[h]
\begin{center}
\includegraphics[width=0.45\textwidth]{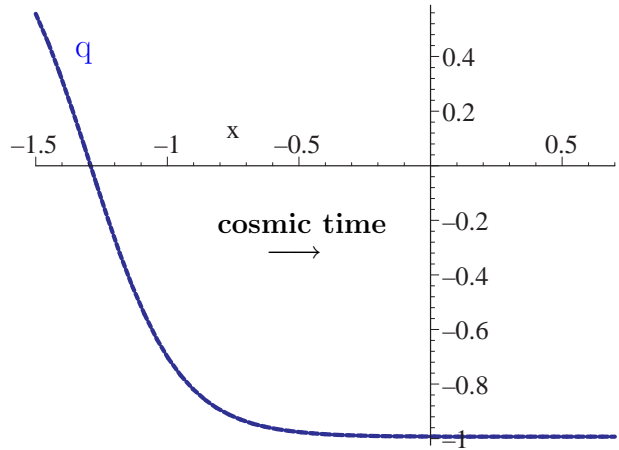}
\end{center}
\caption{The figure shows the deceleration  parameter $q=-\ddot a a/\dot a^2$ against the variable $x$ defined in Eq.~(\ref{def}). The brane is accelerating in the future when $q$ is negative. Notice that $x$ grows as the brane expands and therefore $dx/d\tau>0$ where $\tau$ corresponds to the cosmic time of the brane. This numerical example has been obtained for $b=-1$, $k=1$ and $\beta_1=1$. The last parameter is defined in Eq.~(\ref{def}). Again in order to impose the right initial condition, we started the integration well in the past where CDM dominated over the scalar field on the brane and we can take  as a good approximated solution the dark matter solution given in Eq.~(\ref{asymlambda11}).}
\label{efectomenq}
\end{figure}

Another important question to address is whether the brane is accelerating at the time that the crossing takes place. We know that the vacuum term dominates at late times (see the first assumption). Thus, at that point the brane tension will be adequately described by the vacuum solution; i.e.
\begin{equation}
\bar\lambda\sim C\exp(-x/2) + \ldots\,,\quad C=constant\,>0.
\label{asymlambda21}
\end{equation}
The constant $C$ is positive because for the vacuum solution the brane tension is always positive \cite{Bouhmadi-Lopez:2004ys}. Now, from the results in the previous section, we can conclude that the brane will be speeding up at late times  as long as  $k^2\leq3$. On the other, hand it can be checked numerically that the brane can be accelerating at the crossing as Fig.~\ref{efectomenq} shows.

\section{Conclusions}

In this paper we analyse the behaviour of  dilatonic brane-world models with an induced gravity term on the brane with a constant induced gravity parameter. We assume a  $\mathbb{Z}_2$-symmetry across the brane. The dilatonic potential is an exponential function of the bulk scalar field and the matter content of the brane is coupled to the dilaton field. 
We deduce the modified Friedmann equation for the generalised self-accelerating and generalised normal branch (which specifies the way the brane is embedded in the bulk), the junction condition for the scalar field across the brane and the energy balance on the brane. 

We describe the vacuum solutions; i.e. the matter content of the brane is specified by the brane tension, for a FLRW brane:

\begin{enumerate}

\item In the vacuum self-accelerating branch, the brane tension is a growing function of the scale factor and, consequently, mimics the behaviour of a phantom energy component on the brane. This phantom-like behaviour is obtained without including a phantom fluid on the brane. In fact, the brane tension does not violate the null energy condition. The expansion of the brane is super-inflationary; i.e. the Hubble parameter is a growing function of the cosmic time. At high energy (small scale factors), the brane is asymptotically de Sitter. The brane faces a curvature singularity in its infinite future evolution, where the Hubble parameter, brane tension and scale factor diverge. The singularity happens in an infinite cosmic time. Therefore, the singularity can be interpreted as a ``big rip'' singularity pushed towards an infinite future cosmic time.

\item On the other hand, in the vacuum normal branch, the brane tension is a decreasing function of the scale factor. Unlike the positive branch, the branch is not super-inflating. However, it always undergoes an inflationary expansion (see Fig.~\ref{inflation2}). The inflationary expansion can be eternal ($k^2\leq 3$) or transient ($k^2> 3$), where $k$ is related to the slope of bulk scalar field. For large values of the scale factor, the negative branch is asymptotically Minkowski. 

\end{enumerate}

Furthermore, we have shown the existence of a mechanism that mimics the crossing of the cosmological constant line $w=-1$ in the brane-world scenario introduced in section 2, and which is different from the one introduced in Refs.~\cite{Sahni:2002dx,LDGP2}. More precisely, we have shown that if we consider the  5D dilatonic bulk with an induced gravity term on the normal branch, a brane tension $\lambda$ which depends on the minimally coupled bulk scalar field, and a  brane matter content corresponding only to cold dark matter, then under certain conditions the brane tension grows with the brane scale factor until it reaches a maximum positive value at which it mimics crossing the phantom divide, and then starts decreasing. Most importantly no matter violating  the null energy condition is invoked in our model. Despite the transitory phantom-like behaviour of the brane tension no big rip singularity is hit along the brane evolution (unlike the vacuum self-accelerating branch).

In this  model for the normal branch or non-self-accelerating branch, the constraint equation fulfilled by the brane tension is too complicated to be solved analytically (see Eqs.~(\ref{Friedmann}) and (\ref{constraint})). However, we have shown that under certain physical and mathematical conditions -cold dark matter dominates at higher redshifts and it dilutes a bit faster than dust during the brane expansion as well as a mathematical condition that guarantees the non-existence of a local minimum of the brane tension- it is possible for the brane tension to cross the cosmological constant line. The analytical proof has been confirmed by numerical solutions.  Furthermore, we have shown that for some values of the parameters the normal branch inflates eternally to the future due to the brane tension $\lambda$ playing the role of dark energy through its dependence on the bulk scalar field.

In summary, in the models presented here the mimicry of a phantom-like behaviour and the phantom
divide crossing is based on the interaction between the brane and the
bulk through the brane tension that depends explicitly on the scalar
field that lives in the bulk. We have also shown that in both cases the brane
undergoes a late-time acceleration epoch.

\section*{Acknowledgements}

I am  grateful to Prof P. M. Lavrov and Prof. V. Ya. Epp for a kind invitation to submit this article to the anniversary volume \textit{The Problems of Modern Cosmology}, on the occasion of the 50th birthday of \mbox{Prof. Sergei D. Odintsov}.  I am also grateful to my collaborator A. Ferrera for collaborations upon which some of the work presented here is based.
MBL is  supported by the Portuguese Agency Funda\c{c}\~{a}o para a Ci\^{e}ncia e
Tecnologia through the fellowship SFRH/BPD/26542/2006.

\end{document}